# Chaotic dynamics in spin torque nano oscillator driven by voltage feedback

Meenakshi Sravani[1a)], Swapnil Bhuktare[1]

[1]Department of Electrical Engineering, Indian Institute of Technology Tirupati, Andhra Pradesh, India, 517619

[a)] Author to whom correspondence should be addressed: ee20d503@iittp.ac.in

Non-linear dynamics, including auto-oscillations, chaotic dynamics, and synchronization, are integral to physical and biological applications and can be excited in spintronic devices. In this study, we are interested in exploring the excitation of chaos using voltage feedback in a spin torque nano oscillator using a Magnetic Tunnel Junction (MTJ). According to the Poincaré-Bendixson theorem, chaos cannot arise in a two-dimensional system of MTJ featuring two dynamic variables describing the zenith and azimuth angles of magnetization. Hence, we prefer the feedback system as it creates a multi-dimensional system, making it interesting to explore the emergence of chaos in such systems. Such feedback is achieved by utilizing a 3-terminal device consisting of an MTJ with an in-plane pinned layer (PL) and an out-of-plane free layer (FL) geometry. When a DC current above the critical threshold is applied, the FL's oscillating magnetization generates an AC output voltage through the Tunnel Magneto Resistance (TMR) effect. A fraction of this voltage, fed back after a delay, modulates the FL's anisotropy via voltage controlled magnetic anisotropy (VCMA) effect, potentially driving precessional motion or chaotic dynamics or oscillator death based on the feedback delay and gain of the feedback circuit. The observed chaotic regime has been studied by evaluating the Lyapunov exponent, bifurcation diagrams, Fourier spectral analysis and reconstruction of the trajectory in embedding phase space. Such observed chaotic dynamics can find practical applications in random number generators and physical reservoir computing.

## I. Introduction

The detection and control of magnetization with conventional approaches using voltage or current has streamlined the integration of spintronic devices with the existing CMOS technology [1-2]. Electrical detection of magnetization is characterized by well-known phenomena, namely, Giant Magnetoresistance (GMR) [3-4] in spin valves [5] and Tunnel Magnetoresistance (TMR) [6-7] in Magnetic Tunnel Junctions (MTJ) [8-9]. The control of magnetization by current is achieved by the effect of the Spin Transfer Torque (STT), as demonstrated in prior experimental works [10-11].

Recently, VCMA has become a powerful technique to control magnetization using electric fields [12-13]. It has been reported as one of the energy-efficient methods and exhibits inline properties with high performance (HP) and low power (LP) CMOS in terms of energy-delay products [14]. The origin of this VCMA effect can be attributed to the rearrangement of the atomic orbitals in response to the electric fields near the Ferromagnet (FM)/oxide interface [15]. Additional mechanisms such as charge trapping [16], Rashba coupling [17], and the formation of an electric quadrupole [18] have also been identified as potential origins of the VCMA effect. Using the VCMA effect, people have developed magnetic memory and oscillator-based devices [19-23]. VCMA-based magnetization switching was demonstrated both theoretically and experimentally [19-21]. Spin torque nano oscillators (STNOs) based on this VCMA effect have also been realized [22-23]. So, it is of interest to study the various nonlinear dynamics such as chaos in STNOs using feedback through the VCMA effect.

Chaos is a highly complex yet deterministic dynamic that has a sensitive dependence on the initial set of variables [24-26]. This phenomenon is widely used in practical applications like neuromorphic computing [27-28], random number generators [29-30], and physical reservoir computing [31-32]. The generation of chaotic dynamics has been studied in spintronics using techniques like forcing the system with an external periodic signal [33], devices with non-uniform spatial coupling [34], spin vortex pairs [35], employing magnetic or electric coupling between two ferromagnetic layers [36-37], and delayed feedback mechanisms [38-39]. The macrospin model of STNO features two dynamic variables describing the zenith and azimuth angles of magnetization [40]. According to the Poincaré-Bendixson theorem [41], chaos cannot arise in a two-dimensional system, hence we prefer the feedback system for studying emergence of chaos. Feedback creates a multi-dimensional system, making it interesting to explore the emergence of chaos in such systems. The delayed feedback mechanism based on charge current [42] and magnetic field [43] feedback has been used to excite the chaotic dynamics. The emergence of chaos has also been demonstrated experimentally in vortex-based STNOs with delayed current feedback [43]. The use of voltage feedback using the VCMA effect to excite chaotic dynamics in a STNO has been unexplored. Hence the VCMA effect-based feedback mechanism has been used in this work to study the emergence of chaos.

In this paper, we study the role of VCMA feedback on magnetization dynamics using numerical simulation of the LLGS equation. The chaotic dynamics have been characterized through the study of the Fourier spectrum of the temporal magnetization dynamics and the evaluation of the Lyapunov exponent and bifurcation diagrams using the local maxima of the magnetization oscillations, and phase space trajectory reconstruction [44]. The variation of the Lyapunov exponent with parameters such as feedback gain factor and delay value are also studied. Chaotic dynamics are observed over a wide range of currents and delay values, indicating a broad tunability for chaotic behavior using external parameters. Since the noise limit often reflects the value of the Lyapunov exponent but cannot distinguish between negative and zero Lyapunov exponents, we have preferred to determine chaotic behavior using the Lyapunov exponent rather than the noise limit method [45].

## II. Device structure

The device structure used for the simulations shown in Fig.1 (a), is a 3-terminal device with a stack of Conductor/Insulator/FL/Insulator/PL. MTJ between the terminals T1 and T2 consists of a PL with magnetization lying in-plane along the x-direction and an FL out-of-plane along the z-direction. When a DC current above a critical value ($I_c$) is passed through the MTJ, it exerts STT on the adjacent FM compensating for the damping torque resulting in sustained magnetization oscillations. These oscillations result in AC voltage at the MTJ output due to the TMR effect. The fraction of AC output voltage is then fed back to the conductor terminal T3 with some delay through the delay element. This fed-back voltage modulates the magnetic anisotropy of the FL due to the VCMA effect. Based on the feedback gain factor and the delay the dynamics of the device can be modulated. Since the top MgO between PL and FL is comparatively thinner than the bottom MgO between FL and conductor, the VCMA effect is observed only near the bottom MgO. Please note that similar structures with negative capacitance-enhanced VCMA effect have already been proposed in the literature for oscillator and magnetic memory applications [45-46].

## III. Simulation Methodology

A detailed theoretical study has been done through the macrospin simulations using the well-known LLGS equation (1) [47]. The macrospin approximation is taken due to the small dimensions for practical modelling of the device.

$$\dot{m} = -\gamma m \times H_{eff} - \alpha\gamma\{m \times (m \times H_{eff})\} + H_{STT}\, m \times (p \times m) \quad (1)$$

where

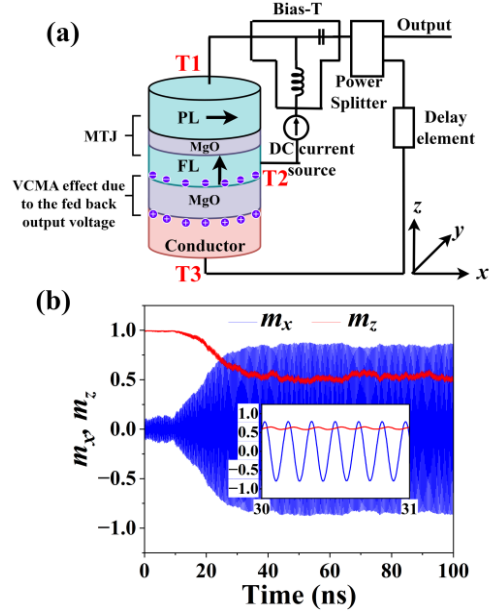

FIG. 1. (a) It is a three terminal device structure consisting of MTJ between two terminals T1 and T2 whereas the other terminal T3 provides feedback to the FL via a conductor and insulator. The output voltage and the input DC current of the MTJ are separated using a Bias-T. A power splitter is used to simultaneously read the output voltage and feed it back to the FL via conductor. Delay is provided using a delay element in the feedback. (b) Magnetization dynamics without feedback for I=1.4 mA.

$$H_{STT} = \frac{\gamma \hbar \eta I}{2q(1 + \lambda m \cdot p)M_S V}$$

$m$ is the normalized magnetization vector, $p$ is the spin polarization vector of the PL, $\gamma$ is the gyromagnetic ratio, $\alpha$ is the Gilbert damping parameter, $\hbar$ is the reduced Planck's constant, $I$ is the current passing through the device, $q$ is the electron charge, $M_s$ is saturation magnetization of the FL, $V$ is the volume of the FL, η is the polarization factor, λ is the dimensionless angular dependency parameter. The first term on the right-hand side of Eq. (1) is the precession term, followed by damping and STT terms.
Effective magnetic field $H_{eff}$ acting on the FL can be given by (2)

$$\boldsymbol{H}_{eff} = (H_k - 4\pi M_S)m_z\hat{\boldsymbol{z}} + H_{ext}\hat{\boldsymbol{z}} + H_{VCMA}m_z\hat{\boldsymbol{z}} + \boldsymbol{H}_{thermal} \quad (2)$$

where $4\pi M_S$ is the demagnetization field, $H_k$ is the perpendicular magnetic anisotropy field, $H_{VCMA}$ is the magnetic field due to the VCMA effect arising out of voltage feedback, $\boldsymbol{H}_{thermal}$ is the random thermal field which is given by Eq. (3)

$$\boldsymbol{H}_{thermal} = \sqrt{\frac{2\alpha kT}{\gamma \mu_0 M_S V \Delta t}} \boldsymbol{G}_r \quad (3)$$

where $k$ denotes the Boltzmann constant, $\mu_0$ stands for the permeability of vacuum, $\Delta t$ signifies the simulation step size, $T$ represents the absolute temperature, and $\boldsymbol{G}_r$ is a random vector drawn from the standard normal distribution.

All the macrospin simulations are run for a duration of 1μs with a step size $\Delta t$ of 1ps. We have performed macrospin simulations by solving the LLGS equation using numerical methods. We employed Heun's method [48] and implemented Stratonovich's calculus approach, utilizing Langevin dynamics [49]. All the parameters are taken from the reference [50-51]: $M_S$ =1448.3 emu/c.c, $\gamma$ = 17.32 MHz/Oe, $H_{ext}$ =2 kOe, $H_k$ =18.6 kOe, $\eta$=0.54, $\lambda = \eta^2$, $V$ = 60×60×π×1.1 nm³, $\alpha$ = 0.005, $\boldsymbol{p} = \hat{\boldsymbol{x}}$, $\beta$= 60 fJ/V-m, bottom MgO thickness $t_{ox}$= 1.4 nm, $\Delta R$= 200Ω. In the absence of the feedback, the critical current is given by Eq. (4)

$$I_c = \frac{4\alpha q M_S V}{\hbar \eta \lambda}(H_{appl} + H_k - 4\pi M_S) \quad (4)$$

The critical current $I_c$ for these parameters to excite the auto oscillations is around 0.8 mA. The magnetization dynamics above this value of critical current for $I$=1.4 mA has been shown in Fig. 1(b)

**Feedback description:**

When DC current $I > I_c$ is passed through the 3-layer MTJ between the terminals T1 and T2, the magnetization of the FL undergoes consistent oscillations. This changes the TMR of the MTJ which is given by $R(t) = R_P + (\Delta R/2)(1 - m_x(t))$ where $\Delta R = R_{AP} - R_P$, $R_P$ is the parallel state resistance and $R_{AP}$ is the antiparallel state resistance. This oscillating resistance gives an ac output voltage $V_{ac}(t) = -I \Delta R m_x(t)/2$ corresponding to the DC current I at terminal T1. The feedback gain fraction (ζ) of this oscillating AC voltage is fed back with some delay $\tau$ and applied to the free layer via conductor (at terminal T3) producing a voltage drop $V_{ac}(t - \tau) = -\zeta I \Delta R m_x(t - \tau)/2$ across the FL. This voltage modulates the PMA field of the free layer due to the VCMA effect. So, the effective VCMA field acting on the free layer due to the feedback is given by

$$H_{VCMA} = -\frac{2\beta V_{ac}(t-\tau)}{\mu_0 M_S t_{ox} t_{FL}} \quad (5)$$

where $\beta$ is the material-dependent VCMA coefficient, $t_{ox}$ is the thickness of the oxide layer, $t_{FL}$ is the thickness of the FL, $V_{ac}$ is given by Eq. (4). The effective magnetic field in the z-direction is modulated by the VCMA field as $(H_k - 4\pi M_s + H_{VCMA})m_z\hat{\boldsymbol{z}}$ affecting the magnetization dynamics of the FL.

### IV. Simulation Results

The macrospin simulations are conducted to explore the various dynamical regimes that emerge using VCMA feedback. As indicated by Eq. (5), the VCMA field is influenced by the control parameters, specifically the feedback gain factor (ζ) and the delay value (τ). The feedback gain factor (ζ) represents the proportion of the AC output voltage that is fed back to the FL via the conductor. To examine how magnetization dynamics vary with ζ, the delay value is fixed at 5 ns—significantly greater than the oscillation period and within an experimentally realizable range. Figure 2 illustrates the magnetization dynamics as ζ varies, with a constant DC current of 1.4 mA. At a lower feedback gain factor of ζ=0.01, the magnetization exhibits small oscillations, as shown in Fig. 2(a). The power spectral density (PSD) of the magnetization $m_x(t)$, obtained from Fourier spectral analysis of the time-series data, indicates that these small amplitude oscillations correspond to a lower PSD with a single peak at a lower frequency, as seen in Fig. 2(b). As the feedback gain factor increases to ζ=0.1, the magnitude of the voltage feedback also rises, modulating the effective magnetic field acting on the FL. This modulation creates an imbalance between the spin-transfer torque (STT) and the damping term, leading to chaotic behavior, which is depicted in the temporal dynamics in Fig. 2(c) and the corresponding Fourier spectrum in Fig. 2(d). In this chaotic regime, the Fourier spectrum shows a wide spread of frequency components, as seen in Fig. 2(d). Further increasing the value of ζ can result in the phenomenon known as oscillator death, illustrated in Figs. 2(e) and 2(f). Oscillator death occurs at high feedback rates when the over-damping of the magnetization leads to a temporarily stable state, which can revert to chaotic behavior when an imbalance between the damping and STT arises. Such phenomena have been observed in coupled oscillators with high feedback rates [52-53]. Since distinguishing chaotic behavior using spectral analysis alone is challenging, additional methods such as Lyapunov exponent estimation, bifurcation diagrams, and the reconstruction of the temporal trajectory in the embedding phase space were employed for further analysis.

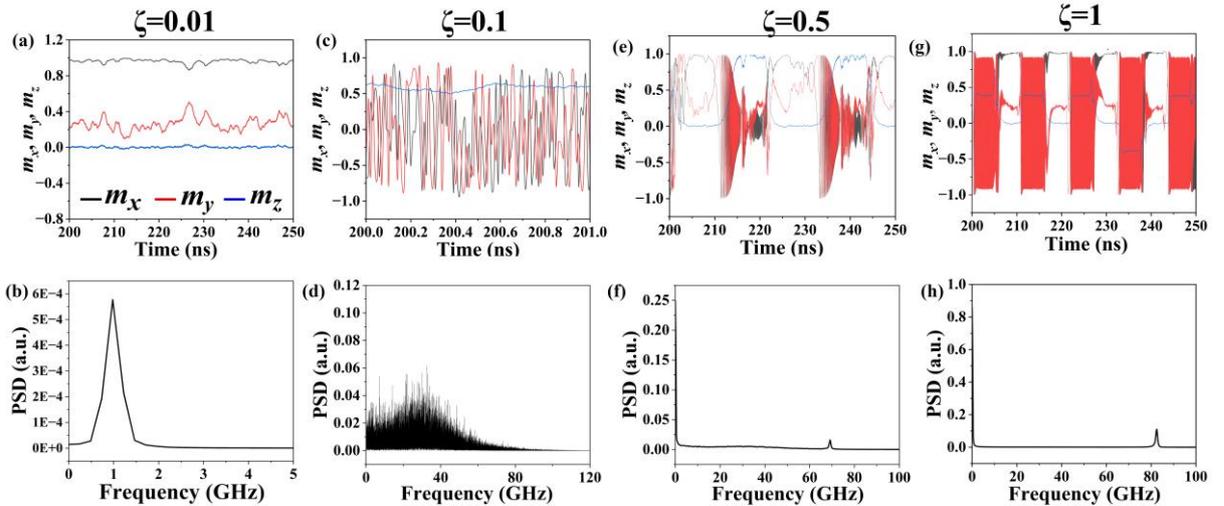

FIG. 2. Temporal magnetization dynamics of the magnetization $m_x, m_y, m_z$ for (a) ζ=0.01 (c) ζ=0.1 (e) ζ=0.5 (g) ζ=1 for a fixed τ= 5 ns, and I = 0.8 mA. Power spectral density of the $m_x$ of the corresponding magnetization for (b) ζ=0.01 (d) ζ=0.1 (f) ζ=0.5 (h) ζ=1. $m_x, m_y, m_z$ colormap shown in Fig. 1(a) has been followed for the remaining plots in Fig. 1(c),1(e), and 1(g).

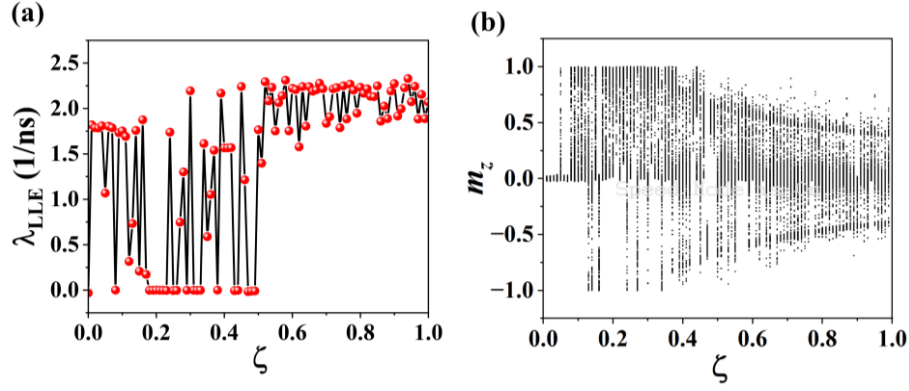

FIG. 3. For a fixed delay of τ= 5 ns, and I = 1.4 mA, (a) Lyapunov exponent $\lambda_{LLE}$ as a function of ζ (b) Local maximum of $m_z$ as a function of ζ.

The chaotic dynamics region has been observed by studying the variation of Lyapunov exponent with ζ. The Lyapunov exponent ($\lambda_{LLE}$) is used to distinguish between the chaotic and non-chaotic regimes based on the sign of the Lyapunov exponent. The positive sign for the Lyapunov exponent classifies the system to be in chaotic regime. The negative sign for the Lyapunov exponent indicates the system to be in an oscillating regime, and a zero value for the Lyapunov exponent corresponds to the system relaxing towards a stable state. In this work, it is estimated using the method which has been specified in the reference [42]. It signifies the mean expansion rate of the two points on a dynamical trajectory which are initially separated by a small distance ε and can be obtained by

$$\lambda_{LLE} = \frac{1}{N}\sum_{i=1}^{N} \lambda_i$$

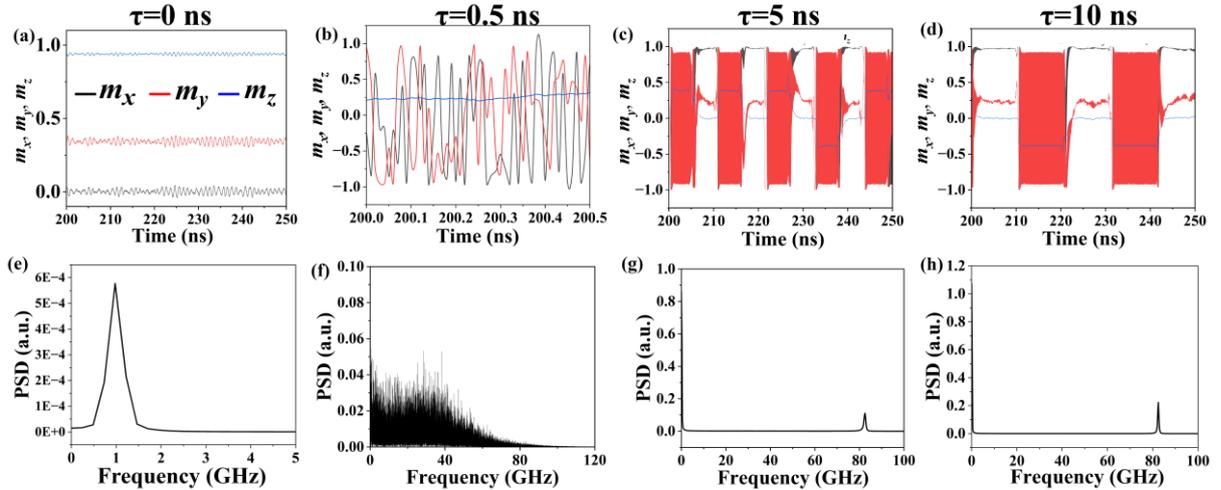

FIG. 4. Temporal magnetization dynamics of the magnetization $m_x, m_y, m_z$ for (a) τ= 0 ns (c) τ= 0.5 ns (e) τ= 5 ns (g) τ= 10 ns for a fixed ζ=1, and I = 1.4 mA. Power spectral density of the $m_x$ of the corresponding magnetization for (b) τ= 0 ns (d) τ= 0.5 ns (f) τ= 5 ns (h) τ= 10 ns. $m_x, m_y, m_z$ colormap shown in Fig. 1(a) has been followed for the remaining plots in Fig. 1(c),1(e), and 1(g).

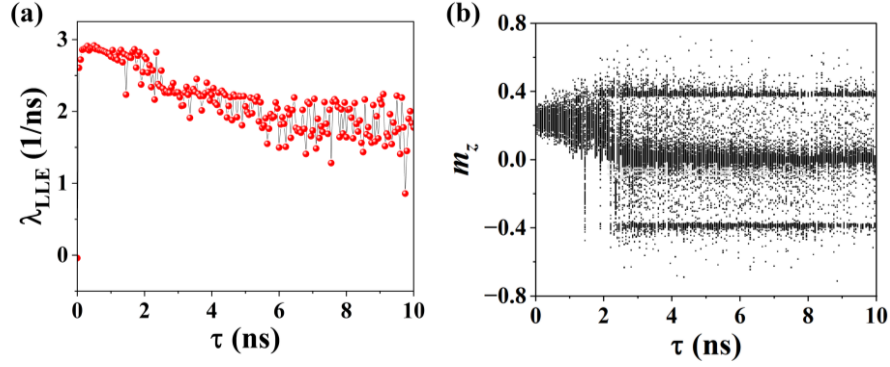

FIG. 5. For a fixed $\zeta=1$, and I = 1.4 mA, (a) Lyapunov exponent $\lambda_{LLE}$ as a function of $\tau$ (b) Local maximum of $m_z$ as a function of $\tau$.

Where $\lambda_i$ is the temporal Lyapunov exponent obtained by shifting **m**(t) to N arbitrary directions and given by $\lambda_i = (1/\Delta t)\ln(\varepsilon_i/\varepsilon)$ where $\varepsilon_i$ is the final separation between the two points on the dynamical trajectory ($i = 1$ to $N$) and $\varepsilon$ is the initial separation.

The Lyapunov exponent has been evaluated as a function of ζ shown in Fig. 3(a). The positive Lyapunov exponent in this figure indicates the region where a chaotic region has been observed. The bifurcation diagram has also been studied by evaluating the local maximum of $m_z$ where a single value indicates the system to be non-chaotic and multiple values indicate the system to be in a chaotic region. Elaborating, a single value for the local maximum occurs when the system is relaxed to a stable state or have a symmetric distribution for the system with small periodic oscillations around a constant value which classifies the system as non-chaotic. Whereas the existence of multiple values for the local maximum can be classified as a chaotic regime. This bifurcation diagram can be seen in Fig. 3(b) as a function of ζ.

A similar kind of analysis has been done to study the variation with delay $\tau$. First, the temporal magnetization dynamics have been shown by varying the $\tau$. To study this variation, we have fixed the value of feedback gain factor $\zeta = 1$. For a delay value of $\tau = 0$, the magnetization shows very small oscillations around a constant value as can be seen in Fig. 4(a), hence corresponding to a single frequency peak as shown in Fig. 4(b). It is important to note that τ=0 does not imply the absence of feedback; rather, it indicates that the output of the MTJ is fed back without any phase shift, meaning feedback is still present. In contrast, when ζ=0, no output is fed back into the system, effectively resulting in a system with no feedback. Now, as the delay is increased ($\tau = 0.5\ ns$) which increases the dimensionality of the system and moves the system into a chaotic state as shown in Fig. 4(c) with the widespread frequency components as shown in Fig. 4(d). Still, a further increase in the delay leads to oscillator death where the system relaxes to a stable state for some time and then jumps back to an oscillating state. An increase in feedback delay corresponds to the increase in the available past information of the magnetization state which leads to an increase in dimensions of the phase space that shows the probability of achieving chaotic behavior This can be seen in Fig. 4(c)-4(d) with their PSD shown in Fig. 4(g)-4(h). It can also be clearly seen from Fig. 4(c)-4(d) that the time of oscillation death is approximately equal to the delay value. The oscillation frequency component at high frequencies is at the same value for both the delay values which is around ~81 GHz as shown in Fig. 4(g)-4(h). Further, for the classification of chaotic and non-chaotic regimes, the Lyapunov exponent and the bifurcation diagram are shown as a function of $\tau$ in Fig. 5(a)-5(b). For only a small delay value, the system is non-chaotic and moves to chaotic regions as the delay is increased.

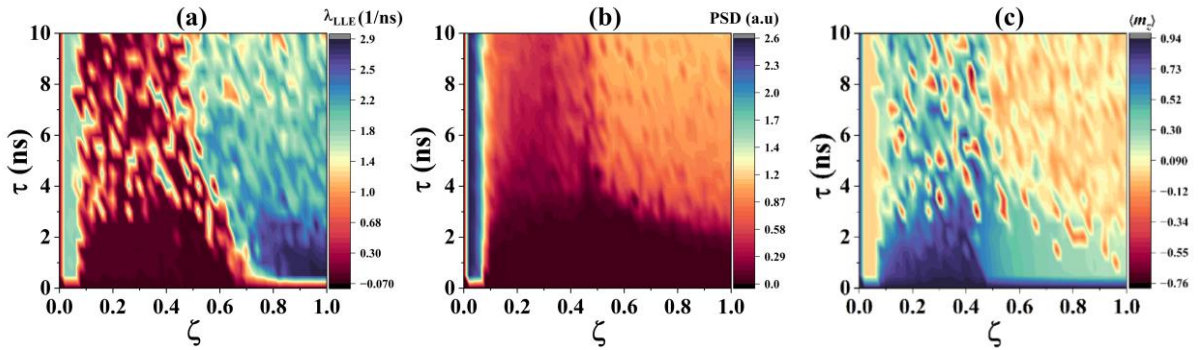

FIG. 6. For a fixed I = 1.4 mA, 2D variation of the (a) Lyapunov exponent as a function of ζ and $\tau$, (b) PSD as a function of ζ and $\tau$ (c) mean of the local maximum of $m_z$ as a function of ζ and $\tau$.

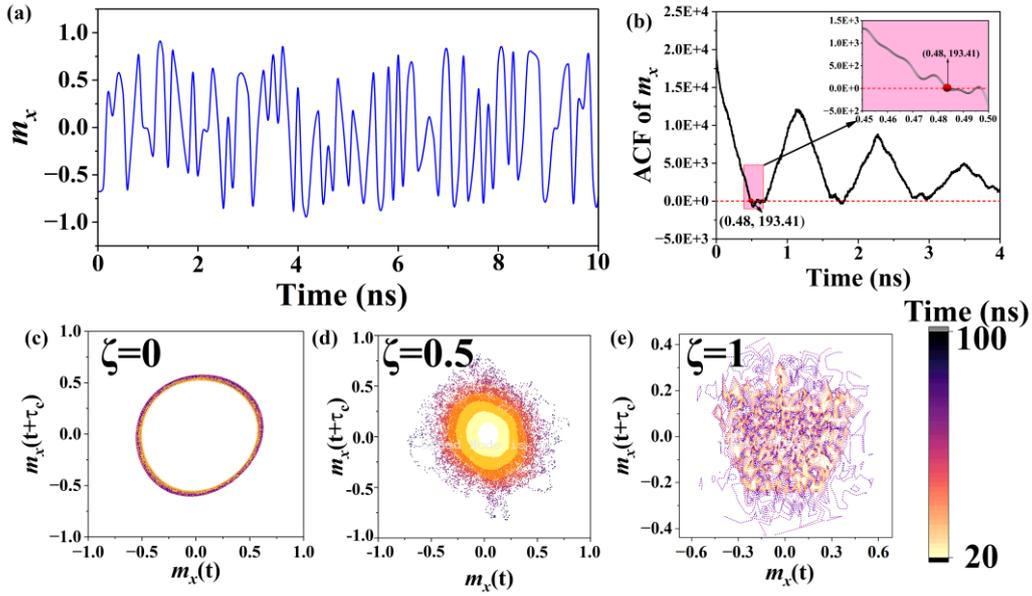

FIG. 7. (a) Time variation of $m_x$ for $\tau = 0.5\ ns$ and $\zeta=1$, (b) The auto correlation function of the $m_x$ with respect to time, Trajectory of the magnetization in embedding phase space for a fixed $\tau = 0.5\ ns$ for (c) $\zeta=0$ (d) $\zeta=0.5$ (e) $\zeta=1$ shown for a duration of 20 to 100ns.

The 2D variation of the Lyapunov exponent and PSD as a function of both $\zeta$ and $\tau$ has been plotted to find the range of values where the system tends to be chaotic. This variation can be seen in Fig. 6(a)-6(c). For lower delay and feedback gain factor, the system is non-chaotic and has a negative Lyapunov exponent with low values of PSD. For higher delay and feedback gain factor values the system moves to the chaotic region with higher PSD and to the region of oscillation death. It is worth noting that the oscillating region within the oscillator death regime is not periodic, and thus the Lyapunov exponent for that region is positive.

Another approach to observe the chaotic region is to reproduce the dynamic trajectory of the magnetization in embedding phase space. This has been done by calculating the minimum time $\tau_m$ at which the autocorrelation function (ACF) of the magnetization $m_x(t)$ goes to zero. For $\zeta = 0$, which implies there is no feedback, the trajectory is circular in real space and can be mapped to a circular trajectory in the phase space. To estimate the minimum time when the ACF of $m_x(t)$ is zero, we have plotted the $m_x$ as a function of time as shown in Fig. 7(a) and its ACF in Fig. 7(b). By repeating the estimation of minimum time $\tau_m$ for several values of $\zeta$, the value is estimated to be around 0.48 ns. Using this $\tau_m$, we have plotted the magnetization trajectory in the embedding phase space for different values of $\zeta$ as shown in Fig. 7(c)-7(e). The trajectory for $\zeta = 0$ is circular, indicating without feedback the system exhibits periodic oscillations. As the $\zeta$ has been increased the trajectory deviates from the circular path and becomes more diversified in the embedding phase space. The temperature effect can also lead to such spread of the phase space. But as can be seen from Fig. 7(c), temperature leads to a slight spread but still the trajectory being circular whereas at higher values it deviates and spreads in the phase space and hence can be identified as chaotic dynamics.

## V. Conclusion

We have employed the feedback mechanism using the voltage control of magnetization through the VCMA effect. This technique led to the emergence of a transition between a stable state, chaotic regime, and oscillation death. Remarkably, the introduction of feedback facilitated the expansion of the phase space to multi-dimensions and hence can be used for the realization of chaotic dynamics within a 2-dimensional system. These chaotic dynamics have been characterized using temporal magnetization dynamics and their PSD, bifurcation diagrams, estimation of the Lyapunov exponent, and reproduction of real space trajectory in embedding phase space. Such chaotic dynamics find potential applications in neuromorphic computing, random number generators, and physical reservoir computing.

### Acknowledgement

We gratefully acknowledge the support received by the Science and Engineering Research Board (SERB), Department of Science and Technology, Government of India through the project EEQ/2020/000164 and CRG/2022/007360.

### Data Availability

The data that support the findings of this study are available from the corresponding authors upon reasonable request.